%% file: sampta13.tex
\documentclass[conference,a4paper]{IEEEtran}
\usepackage{cite}
\usepackage[pdftex]{graphicx}
\usepackage[cmex10]{amsmath}
\usepackage{amssymb}

\usepackage{fixltx2e}

\input{macros}

\begin{document}
%
\title{An Uncertainty Principle for Discrete Signals}

\author{\IEEEauthorblockN{Sangnam Nam}
\IEEEauthorblockA{
Aix Marseille Universit\'e, CNRS, Centrale Marseille, LATP, UMR 7353\\
13453 Marseille, France\\
Email: nam.sangnam@cmi.univ-mrs.fr}}


%


\maketitle

\begin{abstract}
By use of window functions, time-frequency analysis tools like Short Time Fourier Transform
overcome a shortcoming of the Fourier Transform and enable us to study the time-frequency
characteristics of signals which exhibit transient oscillatory behavior.
Since the resulting representations depend on the choice of the window functions, it is 
important to know how they influence the analyses.
One crucial question on a window function is how accurate it permits us to analyze the signals
in the time and frequency domains. 
In the continuous domain (for functions defined on the real line), the limit on the accuracy
is well-established by the Heisenberg's uncertainty principle when the time-frequency spread
is measured in terms of the variance measures.
However, for the finite discrete signals (where we consider the Discrete Fourier Transform),
the uncertainty relation is not as well understood.
Our work fills in some of the gap in the understanding and states uncertainty relation for
a subclass of finite discrete signals.
Interestingly, the result is a close parallel to that of the continuous domain: the time-frequency
spread measure is, in some sense, natural generalization of the variance measure in the continuous
domain, the lower bound for the uncertainty is close to that of the continuous domain,
and the lower bound is achieved approximately by the `discrete Gaussians'.
\end{abstract}


%
\IEEEpeerreviewmaketitle

\section{Introduction}

%

Fourier Transform, due to the fact that it is a global transform, is not
well-suited for the analysis of signals that exhibit transient behavior.
This is a rather significant drawback since such signals exist in
abundance.
One way to remedy this shortcoming is the use of window functions:
a window function enables us to localize the function
to some specific interval of interest that we want to look at.
This gives rise to time-frequency analysis and makes it
possible for us to study the frequency structure of functions at
varying points in time.
Just like Fourier analysis, time-frequency analysis is a fundamental
tool in science, especially in signal processing.

In this article, we define the Fourier Transform $\hat f$ of a complex-valued
function $f$ defined on the real line $\RR$ via
\begin{equation}
 \label{eq:continuousFourier}
 \hat f(\xi) := \int_\RR f(t) e^{-2\pi i \xi t}\, dt, \quad \xi\in\RR.
\end{equation}
The Windowed Fourier Transform of $f$ with a given window function
$g:\RR \to \CC$ would then be defined as
\[
 \gtf_g(\tau, \xi) := \int_\RR f(t) \overline{g(t-\tau)} e^{-2\pi i \xi t}\, dt, \quad \tau, \xi\in\RR.
\]
If $g$ and $\hat g$ are supported near the origin, one may interpret that $\gtf_g f(\tau, \xi)$ is the `$\xi$-frequency
content of $f$ at time $\tau$'.


Unfortunately, such an ideal interpretation cannot become a reality;
the well-known uncertainty principles expresse the idea that 
there is a fundamental limit on how $g$ and $\hat g$ can be simultaneously localized in the two domains.
The most famous formulation of the uncertainty principle is given by the
Heisenberg-Pauli-Weyl inequality (see, e.g., \cite{gr01}):
\begin{thm}
\label{thm:heisenberg}
For $f\in L_2(\RR)$, define the variance of $f$ by
\begin{equation}
\label{eq:varCont}
v_f := \min_{a\in\RR} \frac{1}{\nnorm{f}_2^2} \int_{-\infty}^\infty (t-a)^2 |f(t)|^2 \, dt.
\end{equation}
Then,
\[
 v_f v_{\hat f} \ge \frac{1}{16\pi^2}.
\]
 Equality holds if and only if $f$ is a multiple of $\varphi_{a,b}$,
 defined by
 \[
  \varphi_{a,b}(t) := e^{2\pi i b (t-a)} e^{-\pi (t-a)^2/c}
 \]
 for some $c > 0$.
\end{thm}

%

We may define the mean of $f$ by 
\[
\mu_f := \argmin_{a\in\RR} \frac{1}{\nnorm{f}_2^2} \int_{-\infty}^\infty (t-a)^2 |f(t)|^2 \, dt.
\]
Clearly, the smaller $v_f$ is, the more concentrated the function $f$
is around $\mu_f$.
In other words, $v_f$ is a measure of time-spreading of $f$.
Similarly, $v_{\hat f}$ is a frequency-spreading measure
of $f$.
Thus, the Heisenberg-Pauli-Weyl inequality expresses
the intrinsic limit on how well an $L_2(\RR)$ function can be localized
on the time-frequency plane.
Moreover, the theorem also tells us what the minimizing functions are.

While the Heisenberg Uncertainty Principle gives us a clear picture of 
what can be achieved for time-frequency localization for the continuous
functions defined on $\RR$, our discussion so far is somewhat detached from reality;
we can only consider functions defined on finite intervals in real life.
Furthermore, in this day and age of computers, processing can be done
only when the signal can be stored in memory.
Therefore, the signals are discrete and finite.

A pertinent question is: what can be said about the uncertainty for 
the time-frequency analysis when the Discrete Fourier Transform is used?
Is there any relation between the uncertainties for the continuous and the
discrete cases?
To our knowledge, surprisingly little is known for this problem,
and this is the area that we aim to contribute to with our work.

\section{Discrete Uncertainty Relations: Some Related Works}

In this section, we discuss some works in the literature which may serve
as an introduction to the problem that we are interested in.

\subsection{Uncertainty for Continuous Functions Defined on the Circle}

The Fourier series for periodic functions may be viewed as something
intermediate between the continuous Fourier Transform for functions on
the real line and the discrete Fourier Transform for finite signals.
It could be a good starting point of our discussion on the uncertainty
for discrete signals.

For a $2\pi$-periodic function $f$, the Fourier coefficients for $f$
is defined by
\[
 \hat f(k) := \frac{1}{2\pi} \int_{0}^{2\pi} f(t) e^{-ikt} \,dt,
 \quad k \in \ZZ.
\]

\emph{Remarks on notations: }
For lightness, we will sacrifice the precision and
use the same notation $\hat f$ to mean various different
Fourier Transforms whose meaning will become clear depending on what $f$ is.
Such a convention extends to $\nnorm{\cdot}$ as well.
We also point out that the definition of the continuous Fourier Transform $\hat f$ 
used in this subsection is defined without the $2\pi$-factor in \eqref{eq:continuousFourier}.


The question we are interested in is how concentrated, or conversely how
spread, $f$ and $\hat f$ are.
We note that even though $\hat f$ is a discrete sequence, there is no problem in
defining the variance of it; we need only to replace the integral in \eqref{eq:varCont}
with an analogous sum. The mean $\mu_{\hat f}$ can be similarly defined.
The situation is different for $f$. 
The issue is that we cannot simply compute
\[
\frac{1}{\nnorm{f}_2^2} \int_{0}^{2\pi} t |f(t)|^2 \, dt
\]
for the mean of $f$. 
Such a quantity fails to take the periodicity into account.

A different way to characterize
`the mean value' had been proposed (see \cite{Breitenberger83uncertainty}):
\[
 \tau(f) := \frac{1}{\nnorm{f}_2^2} \int_0^{2\pi} e^{it} |f(t)|^2 \, dt.
\]
The periodicity is clearly reflected in $\tau(f)$.
With that, one defines `the variance' of $f$ as
\[
 \frac{1}{\nnorm{f}_2^2} \int_0^{2\pi} |e^{it} - \tau(f)| |f(t)|^2 \, dt
 = 1 - \tau(f)^2.
\]
With these time-frequency spread measures, the uncertainty relation for the
continuous functions on the circle was shown to be as follows:
\begin{equation} \label{eq:preUncertaintySemi}
 \left(1-\tau(f)^2\right) v_{\hat f} \ge \frac{\tau(f)^2 }{4}.
\end{equation}
Note that unlike in the continuous setting, the quantity on the right-hand
side depends on the function $f$.
Therefore, if we were to use $(1-\tau(f)^2) v_{\hat f}$ as the measure of
uncertainty of $f$, the equality in \eqref{eq:preUncertaintySemi}
does not immediately imply that $f$ is a minimizer of the uncertainty.
A simple way to bypass this issue is to define the time spread of $f$ as
\[
 v_{f} := \frac{1-\tau(f)^2}{\tau(f)^2}.
\]
A more precise description of the resulting uncertainty principle is given as follows
\cite{Narcowich96wavelets,Prestin99optimal}:
\begin{thm}
 For a function $f \in AC_{2\pi}$ with $f' \in L_2([0,2\pi])$
 where $f$ is not of the form $c e^{ikt}$ for any $c\in\CC$,
 $k\in \ZZ$, it holds that
 \[
  v_f v_{\hat f} > \frac{1}{4}.
 \]
The lower bound is not attained by any function, but is best
possible. Here, $AC_{2\pi}$ is the class of $2\pi$-periodic absolutely continuous functions.
\end{thm}

One reservation towards this result is that the meaning of the
so-called \emph{angular spread} $v_f$ is not very intuitive.
In addition, the theorem does not give any guide on what functions may
have the uncertainty product close to the lower bound.

A result in \cite{prestin03} sheds some light on the second problem.
The authors used a process of periodization and dilation to show that
a sequence of functions achieve the uncertainty for functions defined
on the real line in the limit.
They proved:
\begin{thm}
 For an admissible function $f$ (defined on the real line),
 \[
  \lim_{a\to\infty} \frac{1}{a^2} v_{f_a} = v_f,
  \quad
  \lim_{a\to\infty} a^2 v_{\hat f_a} = v_{\hat f},
 \]
 where
 \[
  f_a(t) := \sqrt{a} \sum_{k\in\ZZ} f(a(t+2\pi k)).
 \]
 Therefore,
 \[
  \lim_{a\to\infty} v_{f_a} v_{\hat f_a} = v_f v_{\hat f}.
 \]
\end{thm}
\
We remind the reader that the definitions of $v_{f_a}$ and $v_f$
are quite different. 

Since the minimum of $v_f v_{\hat f}$ is known to be $1/4$
and is achieved by (essentially) Gaussian functions, 
the theorem provides a way to build periodic functions that
are asymptotically optimal in the given measure of time-frequency
spreads.
We will see that our result shares some similarity with this.

Another way to obtain periodic functions which nearly achieve the
uncertainty bound is by computing directly with numerical optimization
\cite{parhizkar2013sequences}.
In this approach, Parhizkar et al. fixed the angular spread at a
prescribed level and then searched for functions that minimize
the frequency spread with the given angular spread.
They formulated the problem as a quadratically constrained quadratic
program and hence enabled efficient computations of desired window
functions.

For more results for uncertainties for functions on the circle
which include different spread measures, refer to \cite{Ishii:1986uncertainty,calvez:1992uncer,venkateshUncertainty}.

\subsection{Sparsity and Entropy}

There are several works in the literature on the uncertainty relation for 
finite discrete signals where the Discrete Fourier Transform is considered;
see, e.g., \cite{Donoho89uncertainty,Donoho01uncertaintyprinciples,Tao_anuncertainty,meshulam:uncert,GhJa:uncert,
krahmer-pfander-rashkov-08,Przebinda99using}.
Most results in these can be generically stated as $\phi(\x) + \phi(\hat\x) \ge c_s$ or $\phi(\x)\phi(\hat\x) \ge c_p$ for some
constants $c_s$ and $c_p$ where $\phi(\x)$ measures the spread of $\x$.
In \cite{Donoho89uncertainty,Donoho01uncertaintyprinciples,Tao_anuncertainty,meshulam:uncert},
$\phi(\x)$ is chosen to be $\norm{\x}_0$, i.e., the sparsity or the number of non-zero entries of $\x$.
In \cite{Przebinda99using}, the entropy of $\x$, $\Ent(\x)$, is used for $\phi(\x)$.
For more on these and other topics regarding uncertainty principle, 
refer to \cite{ricaud12survey}.

While these results are deep and important with much impact, we note that $\norm{\x}_0$ and $\Ent(\x)$
(and other similar measures) do not reflect properly the underlying geometry.
For example, if $\x$ consists of two pulses, $\norm{\x}_0 = 2$ no matter where the pulses are. 
However, in many contexts, we clearly regard $\x$ is more localized/concentrated if the pulses are next to
each other.

Another potential drawback is that
the minimizers of these uncertainty measures tend to be the picket-fence signals
(Dirac comb).
From the perspective of window signals, those are intuitively regarded
as poorly localized on the time-frequency plane.
These are the reasons why we insist on the definitions in Section \ref{sec:measureUnc}.

Before closing the section, we mention the work \cite{gomi:uncert}. In this work, they
consider two operators (which may not even be self-adjoint) in a Hilbert space and
derive related uncertainty relations. Since their result is general, one can apply
it in the setting that we are interested in and obtain some uncertainty relation.
For appropriate choice of operators, one may obtain a result that would be close to
ours. While interesting, we think this is not a simple task. We also point out that
our result links uncertainty relations in two different domains, which is not addressed
by \cite{gomi:uncert}.

\section{Connection between Discrete and Continuous Uncertainty Relations}

In this section, we present the main result of this paper. 

\subsection{Discretized Time-Frequency Spreads Measures}
\label{sec:measureUnc}

Let us fix a positive integer $N$ and consider the space $\CC^N$ of $N$-dimensional signals.
For our purposes, we will regard a vector $\x\in\CC^N$ as defined on
$N$ uniformly spaced points
\[
 \ddom := \Big\{ -\frac{N}{2}+1, -\frac{N}{2}+1,
 \ldots, \frac{N}{2} \Big\}/\sqrt N.
\]
With this understanding, the Discrete Fourier Transform $\hat \x\in\CC^N$ of $\x\in\CC^N$
is defined by
\[
 \hat\x(k) := \frac{1}{\sqrt N} \sum_{j\in\ddom} \x(j) e^{-2\pi jk}, \quad k \in \ddom.
\]
The inverse transform has the following form:
\[
 \x(j) = \frac{1}{\sqrt N} \sum_{k\in\ddom} \hat\x(k) e^{2\pi jk}, \quad j \in \ddom.
\]

Next, we consider a measure of spread of a vector $\x\in\CC^N$. 
For this, we go back to \eqref{eq:varCont} and adapt it to our setting.
Viewing $|t-a|$ as the distance between $t$ and $a$, it is natural to
define the variance $v_\x$ of $\x\in\CC^N$ by
\[
 v_\x := \min_{a\in \mathcal{I}_N} \frac{1}{\norm{\x}_2^2} \sum_{j\in\ddom} d(j,a)^2 |\x(j)|^2
\]
where $\mathcal{I}_N$ denotes interval $(-\sqrt N, \sqrt N]/2$ and $d(j,a)$ is the distance between $j$ and $a$.
Now note that our definition of Discrete Fourier Transform assumes that the signals
in $\CC^N$ are $\sqrt N$-periodic. Taking this into account, we define
the distance between two points $j$ and $a$ by
\[
 d(j, a) := \min_{l\in\sqrt N\ZZ} |j-a-l|.
\]
Finally, we may define the mean $\mu_\x$ of $\x$ to be the minimizing value $a\in\mathcal{I}_N$
of the right-hand side expression above for $v_\x$.
Note that $v_{\hat\x}$ is identically defined.

\subsection{No Uncertainty?}
\label{sec:noUncertainty}

With our definition of uncertainty $v_\x v_{\hat \x}$, 
there cannot be any uncertainty principle in the conventional sense.
Clearly, for any $\x\in\CC^N$, we have
 $v_\x \le N/4$ and $v_{\hat\x} \le N/4$.
On the other hand, the vector $\x$ that is supported at the origin satisfies $v_\x = 0$.
Hence, $v_\x v_{\hat \x} = 0$.
It appears that there is no uncertainty at all and that we can do as well as
we want!

Of course, such a claim is non-sense, and it runs counter to our intuition that we could not
have a signal localized simultaneously in both domains as accurate as we wanted.
A closer look at the case $v_\x v_{\hat\x} = 0$ reveals why we came to this conclusion.
The signal $\hat\x$ is \emph{globally} supported but $v_{\hat\x}$ fails to express 
the badness in frequency localization since it is always bounded above by $N/4$.
In contrast, one would have had $v_{\hat f} = \infty$ in such cases.
One way to resolve this issue would be to re-define $v_{\hat\x}$ (and $v_{\x}$) in a way
so that $v_{\hat\x} = \infty$ in this kind of signals $\x$.
However, we will not take this route since the argument in Section \ref{sec:measureUnc}
shows that $v_\x$ is a sensible way to gauge the time spread of $\x$.
How can we formulate a sensible uncertainty principle then?

\subsection{Uncertainty for a Subclass of Discrete Signals}

As seen in \ref{sec:noUncertainty}, there are signals that we clearly want to exclude from
our consideration. Thus, it makes sense to
restrict our attention to a subclass of signals in $\CC^N$ in order to exclude
the cases where $\x$ or $\hat\x$ are `globally supported'.

Based on the similarity between the discrete and the continuous Fourier Transforms,
it is natural to suspect that discrete finite samples of Gaussian functions
might be optimal windows for the Discrete Fourier Transform.
While this appears reasonable, it looks difficult to show its validity rigorously.
Moreover and perhaps obviously, taking discrete finite samples of Gaussian functions
would be a bad idea \emph{unless} they happen to be nearly zero outside the
sampling interval.
This leads us to introduce `admissible functions' for our discussion.

We say that 
$f\in L_2(\RR)$ is $(N,\epsilon)$-\emph{localized} if
\begin{equation}
 |f(t)| \le \frac{\epsilon}{|t|^2}, \quad |t| \ge \frac{\sqrt N}{2},
\end{equation}
and that a signal
$\x\in \CC^N$ is \emph{admissible} with constant $\epsilon$ if
\[
 \x(j) = \x_f(j) := N^{-1/4} \sum_{l\in\sqrt N\ZZ} f(j + l), \quad j\in\ddom
\]
for a function $f$ with $(N,\epsilon)$-localized functions $f$, $f'$, $\hat f$, $\hat f'$.
That is, admissible vectors in $\CC^N$ are obtained by uniformly
sampling $\sqrt N$-periodized localized functions in $L_2(\RR)$.

Our main result is the following:
\begin{mthm}
 \label{thm:main}
 Suppose that $f\in L_2(\RR)$ is localized in time-frequency domain with constant $\epsilon$.
 Then,
 \[
  \sqrt{v_f v_{\hat f}} ( 1 - \sqrt\epsilon) \le \sqrt{v_\x v_{\hat \x}}
  \le \sqrt{v_f v_{\hat f}} ( 1 + \sqrt\epsilon),
 \]
 where $\x := \x_f$.
 Thus, if $\x$ is an admissible signal, then
 \[
  v_\x v_{\hat\x} \ge \frac{(1-\sqrt\epsilon)^2}{16\pi^2}.  
 \]
\end{mthm}

To give some idea of the proof, we ask first: Why do we associate $\x_f$ to $f$ instead of sampling the function
directly without periodizing it?
Upon some reflection, the periodization seems to be natural given the well-known
phenomenon of folding (aliasing) associated with sampling approach.
It is the periodization that makes the two endpoints of $\ddom$ to be neighbors
when the sampling is done.
Another crucial reason for us to introduce $\x_f$ in that way is the observation
that $\hat\x_f = \x_{\hat f}$, which is a standard consequence of Poisson Summation Formula.
Thanks to this identity, we need only to show that $v_\x$ and $v_{\hat\x}$ are good approximations of
$v_f$ and $v_{\hat f}$, respectively.
To show that $v_\x$ and $v_f$ are close to each other, we show that relevant moments of $\x$
and $f$ are very close. For this purpose, we apply the Poisson Summation Formula
and the Parseval's identity. This is also where we use $(N,\epsilon)$-localizedness of $f$, $f'$, $\hat f$, and
$\hat f'$. A detailed proof of Theorem \ref{thm:main} will be given in an up-coming work.



\section{Discussion and Conclusion}

One implication of Theorem \ref{thm:main} is that, if we were to consider only the admissible
signals in $\CC^N$ as windows -- which is not unreasonable in many applications since one would
like to have `smooth' and `fast-decaying' windows for the time-frequency analysis -- thanks to Theorem \ref{thm:heisenberg},
we can easily construct nearly optimal windows for the Discrete Fourier Transforms
by periodizing Gaussian functions and taking finite uniform
samples as long as the Gaussian functions are supported essentially on the interval
of sampling. 
This is a mild requirement due to the exponential decay of the Gaussian functions, especially when $N$ is large.

We must keep in mind that `discrete gaussians' above are,
a priori, nearly optimal only among admissible signals in $\CC^N$;
however, we will demonstrate in the up-coming work that
the near optimality of the discrete gaussians may be valid for `all signals' in $\CC^N$.
More theoretical evidence 
related to the near optimality of the discrete Gaussians
will be given there.
We also show by numerical computation that the uncertainty bound 
is indeed very close to $1/(16\pi^2)$.

To conclude, we asserted that the uncertainty products of admissible signals with constant $\epsilon$
in $\CC^N$ are bounded below by constant close to $1/(16\pi^2)$.
Based on this claim, we derived that the discrete Gaussians are near optimal windows
among the admissible signals.

Even though the near optimality of the discrete Gaussians among \emph{all} signals is strongly
suspected, a definitive proof is still missing and remains as future work.
Also, as a side problem, it would be interesting to study the characteristics of the discrete
Gaussians that arise from Gaussian functions with wide support. For example, are those signals
near optimal in some other sense?

Finally, we mention the question of optimal windows for distinguishing, e.g., linear chirps.
In our follow-up, we take the approach of this work and try to establish, at least
formally, that modulated discrete Gaussians (so that they themselves are linear chirps) are
nearly optimal as well.

\section*{Acknowledgment}

This work is supported by the European project UNLocX (FET-OPEN, grant number 255931).
The author would like to thank Bruno Torr\'esani and Benjamin Ricaud for helpful discussions
on the subject. The author also thanks the reviewers for their constructive comments.

\nocite{Folland97uncertainty}


\bibliographystyle{IEEEtran}
\bibliography{refs}
%

%

\end{document}

%% file: macros.tex
\newcommand{\RR}{{\mathbb{R}}}
\newcommand{\CC}{{\mathbb{C}}}
\newcommand{\ZZ}{{\mathbb{Z}}}

\newcommand{\x}{\mathbf{x}}
\newcommand{\norm}[1]{\left\| #1 \right\|}
\newcommand{\nnorm}[1]{\| #1 \|}

\newcommand{\gtf}{{\mathcal{V}}}
\newcommand{\Ent}{{\mathcal{S}}}
\newcommand{\ddom}{{\mathcal{D}_N}}
\DeclareMathOperator*{\argmin}{argmin}

\newtheorem{thm}{Theorem}
\newtheorem{mthm}[thm]{Theorem}